%% file: main.tex
\documentclass[lettersize,journal]{IEEEtran}
\usepackage{amsmath,amsfonts}
\usepackage{algorithmic}
\usepackage{algorithm}
\usepackage{array}
\usepackage{makecell} 
\usepackage[caption=false,font=normalsize,labelfont=sf,textfont=sf]{subfig}
\usepackage{textcomp}
\usepackage{stfloats}
\usepackage{url}
\usepackage{verbatim}
\usepackage{booktabs}
\usepackage{multirow}
\usepackage{graphicx}
\usepackage{cite}
\usepackage{hyperref}   
\usepackage[table]{xcolor} 
\usepackage{pifont}         

\usepackage{amssymb}

\begin{document}

\title{HapticLDM: A Diffusion Model for Text-to-Vibrotactile Generation}

\author{
Jiahao Xiong$^\ast$,
Fei Wang$^\ast$$^\#$,
Anran Xu,
Pinzhi Huang,
Tao Wen,
Lijia Pan,
and Cai Chen$^\#$%
\thanks{This work was supported by Guangdong Provincial R\&D Program in Key Areas under Grant 2023B0101200011.}

\thanks{Jiahao Xiong and Fei Wang contributed equally to this work.}

\thanks{Jiahao Xiong, Fei Wang and Cai Chen are with Technology Development Center, Guangzhou Shiyuan Electronic Technology Company Limited, Guangzhou 510535, China.}

\thanks{Anran Xu and Tao Wen are with School of Automation and Intelligence, Beijing Jiaotong  University, Beijing 100044, China.}

\thanks{Pinzhi Huang is with Department of Architecture, University of California, Berkeley, Berkeley, CA 94720, USA}

\thanks{Lijia Pan is with School of Electronic Science and Engineering, Nanjing University, Nanjing 210093, China.}

\thanks{Corresponding authors: Fei Wang (wangfei8398@cvte.com) and Cai Chen (chencai@cvte.com).}
}



\maketitle

\input{0_abstract}
\input{1_intro}
\input{2_related}
\input{3_method}
\input{4_experiments}

\input{5_results}

\input{6_disscusion}
\input{7_Conclusion}

\input{8_acknowledgement}

\bibliographystyle{IEEEtran}
\bibliography{refs}

\clearpage

\appendices
\input{9_appendix}

\vfill

\end{document}

%% file: 0_abstract.tex
\begin{abstract}
Text-to-vibration generation converts natural language into haptic feedback, enabling vibration-effect designers to get scenarios-fitted vibrations more efficiently, which shows great potentials in application fields such as metaverse, games, and film to enrich the user experience in interactive scenarios. The core challenge in this field is how to generate accurate, consistent, and complete vibrations according to textual semantics. Very recent autoregressive (AR) approaches (e.g., HapticGen) exhibit limited capacity in fully capturing global dependencies, owing to the inherent sequential nature of their modeling and prevailing data constraints. In this paper, we proposed HapticLDM, the first text-to-vibration generative model built upon Latent Diffusion Models (LDMs). Firstly, with respect to the data, we introduced a text-processing strategy that emphasizes dynamic characteristics to curate high-quality data pairs for fine-grained dynamic modeling. Secondly, HapticLDM incorporates a global denoising mechanism that regulates coherent and stable variations in the temporal envelope. Furthermore, we conduct extensive evaluations, including A/B testing against the state-of-the-art baseline and a user study involving 30 participants. The results demonstrate that our model enhances realism and semantic alignment. Qualitative feedback further indicates that HapticLDM simplifies the haptic design workflow while generating diverse, subtle, and physically precise vibrations.
\end{abstract}


%% file: 1_intro.tex
\input{figs/hapticldm_overview}

\section{Introduction}

Vibrotactile feedback plays a critical role in immersive user experiences, enhancing engagement in virtual reality (VR), gaming, films, and assistive technologies \cite{degraen2021weirding,kim2020defining,singhal2021juicy,yun2023generating}. However, designing high-fidelity haptic effects remains a complex and labor-intensive endeavor. It demands specialized expertise, extensive practical experience and iterative refinement \cite{schneider2017haptic,seifi2020novice}, resulting in a high entry barrier for novice designers and limited design efficiency.

Traditional design workflows predominantly rely on manual editing tools to create tactile effects by adjusting parameters such as amplitude, temporal intervals, and frequency \cite{john2024adaptics,seifi2023feellustrator,swindells2006role}. Commercial platforms, such as AAC Technologies’ RichTap Studio \cite{richtap_aac_2026}, provide parameterized control over signal properties, including amplitude envelopes and spectral components — facilitating iterative composition of vibrotactile patterns. While these toolkits are functional, these approaches are largely experience-driven and require substantial expertise \cite{schneider2017haptic}. Existing libraries of predefined vibration patterns further support design through reuse and modification \cite{schneider2016studying,seifi2015vibviz}. However, prior research underscores significant limitations: design creativity is constrained by tool capabilities, and producing high-quality haptic signals still depends on expert knowledge, intuition, and exhaustive trials and errors \cite{schneider2017haptic,maclean2017multisensory,seifi2023feellustrator}.

To mitigate these barriers, recent research has increasingly focused on cross-modal transformation and generation. Given the inherent perceptual coupling between auditory and tactile senses, audio has long been used as a primary proxy for haptic design. Okazaki et al. \cite{okazaki2015effect} demonstrated that frequency-shifting techniques can enhance musical experiences via audio-to-tactile conversion. More recently, Zhan et al. \cite{zhan2023method} proposed a Residual U-Net for audio-to-tactile generation. In industry, Meta introduced audio-to-vibration functionality in Meta Haptic Studio \cite{meta_haptics_studio}, while Sony leveraged audio-driven processing to deliver high-fidelity haptic feedback in the DualSense controller \cite{sony2020haptics_rd}. However, despite their efficacy, these approaches are fundamentally constrained by their reliance on pre-existing audio inputs, lacking the capability to directly generate haptic signals from high-level textual descriptions.

HapticGen \cite{sung2025hapticgen} addresses this limitation by introducing text-to-vibration generation using a generative model. It enables natural language-driven haptic content creation, significantly improving accessibility and efficiency. Its key contributions include constructing a large-scale haptic dataset from audio sources and employing an Autoregressive (AR) model for waveform generation.

However, limitations remain in both data construction and model design. HapticGen builds its dataset using Large Language Model (LLM)-based haptic label augmentation and audio-to-haptic conversion. Since LLM-generated descriptions are guided by linguistic priors rather than signal-level temporal characteristics \cite{huang2025survey}, inconsistencies may arise between textual annotations and actual vibration patterns. Such discrepancies, potentially caused by semantic hallucination \cite{huang2025survey}, introduce noise during training.

From a modeling perspective, AR generation is inherently constrained by its sequential prediction mechanism \cite{kreuk2022audiogen}. Neuroscience studies indicate that perception of mid-to high frequency vibrations (40–400 Hz) is governed primarily by temporal structure encoding rather than instantaneous amplitude \cite{harvey2013multiplexing,weber2013spatial}. Thus, temporal dynamics—such as rhythmic patterns and envelope variations—serve as the primary carriers of haptic semantics \cite{hollins2000evidence}. However, AR models lack explicit mechanisms to capture such global temporal structures, making it difficult to generate coherent long-range patterns. Consequently, generated signals tend to be repetitive or overly uniform \cite{liu2023audioldm}. Furthermore, since each prediction depends on previously generated outputs, errors accumulate over time \cite{ren2020fastspeech2}, further degrading modal alignment. Therefore, both the data construction process and the generative model remain inadequate to achieve effective modal alignment, required in cross-modal generation\cite{li2021align,radford2021learning,wu2023large}.

To improve cross-modal alignment in text-to-vibration generation, we propose the Haptic Latent Diffusion Model (HapticLDM) (Fig. \ref{fig:hapticldm-overview}). First, to mitigate mismatches between converted signals and textual descriptions, we develop an automated data processing pipeline that emphasizes tactile-relevant characteristics and filters out invalid samples. Compared with the LLaMA-3-8B model used in HapticGen \cite{grattafiori2024llama}, we adopt the larger Qwen2.5-32B model \cite{qwen2.5} to enable more accurate extraction and reconstruction of semantics related to temporal dynamics from the raw text. This process removes irrelevant content and rewrites subjective expressions into neutral, vibration-oriented descriptions. As a result, temporal structures can be represented more precisely, improving cross-modal consistency.

Second, to better model temporal structure, we introduce Latent Diffusion Models (LDMs) for vibrotactile generation, allowing for the generation of 10-second vibration sequences directly from textual input. To the best of our knowledge, this is the first application of LDMs to haptic signal generation. Our approach adapts the Stable Audio architecture \cite{evans2025stable} to the haptic domain. The global denoising process in latent space aligns naturally with the dependence of vibration signals on global envelope evolution. By modeling signal envelopes holistically, rather than relying on point-wise autoregressive prediction, HapticLDM achieves stronger temporal coherence and physical continuity over long sequences. We evaluate our method using VR controllers from Meta Quest 3 \cite{meta_quest3_2023}, conducting both user studies and A/B testing on this platform.

In summary, the contributions of this work are as follows:
\begin{itemize}
\item \textbf{Dataset Construction:} We propose an automated data processing pipeline that enhances tactile-relevant characteristics in the dataset. Through filtering and objective rewriting, textual descriptions are transformed into neutral, vibration-oriented representations, reducing cross-modal inconsistencies and improving training quality.
\item \textbf{Generative Method:} We introduce HapticLDM, the first latent diffusion framework for haptic signal generation. The model outperforms in global temporal structure generation. Its superiority in immersive haptic generation was also demonstrated in following.
\end{itemize}

%% file: figs/hapticldm_overview.tex
\begin{figure*}[ht] 
    \centering
    \includegraphics[width=\linewidth]{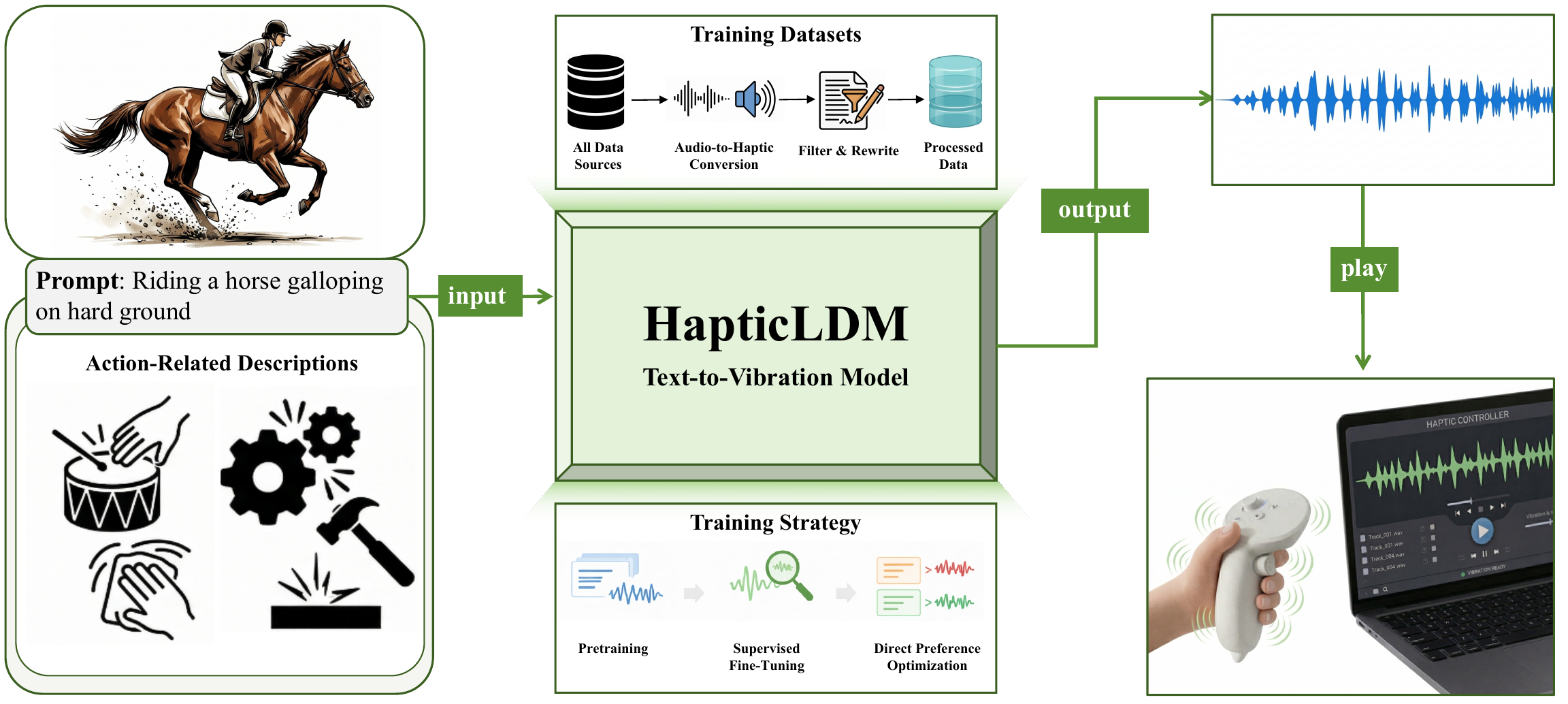}
    \caption{Overview of HapticLDM. A text prompt is initially provided as input. A corresponding vibration signal is then generated using HapticLDM. Output vibrations are displayed on a desktop-based user interface. Finally, generated vibrations are evaluated by designers through playback on VR controllers.
    }
    \label{fig:hapticldm-overview}
\end{figure*}

%% file: 2_related.tex
\section{Related Work}\label{sec:related}

\subsection{Haptic Dataset Development and Augmentation}

Compared with text-to-image and text-to-audio generation, text-to-vibration generation is constrained by the lack of large-scale datasets \cite{sung2025hapticgen}. Existing vibration datasets are limited in scale, annotation richness, and task relevance. Early vibrotactile libraries, such as VibViz \cite{seifi2015vibviz} and Feel Effects \cite{israr2014feel}, provide only a small number of signals with tags or ratings. While useful for browsing and perceptual analysis, they are insufficient for training modern text-conditioned generative models. Many datasets also focus on textures and material properties, such as the LMT haptic texture database \cite{strese2014haptic} and the Penn Haptic Texture Toolkit \cite{culbertson2014modeling}. These datasets cover only a limited range of vibrotactile patterns required in interactive applications. In addition, RecHap \cite{theivendran2023rechap} augments a small set of mid-air ultrasound haptic signals through geometric transformations. However, this approach targets a different haptic modality and provides limited semantic diversity for text-to-vibration tasks.

Recent work has begun to address the lack of textual annotations for vibration signals. HapticCap \cite{hu2025hapticcap} introduced the first large-scale, fully human-annotated vibration-caption dataset, significantly enriching the language component of haptic data. However, it is primarily designed for caption retrieval and understanding rather than generation. Moreover, constructing such datasets is time-consuming and costly, as users must physically experience each signal before annotation. This process is further complicated by the lack of standardized vocabulary for vibration description \cite{obrist2013talking} and the difficulty of interpreting haptic signals without contextual cues \cite{maclean2008foundations}.

To address these challenges, HapticGen \cite{sung2025hapticgen} proposed a cross-modal bootstrapping strategy. Instead of relying on a large text-vibration corpus, it converts the captioned audio dataset WavCaps \cite{mei2024wavcaps} into vibration signals and augments captions with haptic-oriented descriptions using an LLM \cite{grattafiori2024llama}. This approach effectively expands the training data and enables large-scale text-to-vibration generation. HapticGen also introduces expert- and user-preference datasets to improve alignment. However, the augmented texts are still derived from audio captions and LLM-generated descriptions, rather than grounded in the actual temporal dynamics of vibration signals. As a result, textual descriptions may include temporal details that are not reflected in the corresponding signals. For example, when HapticGen utilizes an LLM for text augmentation with the system prompt, ``Write a tactile expression description sentence in third person perspective action'', an original audio caption such as ``A thunderstorm is happening." is expanded into: ``The air vibrates with intense tremors as the thunderstorm rages on, the ground shuddering beneath her feet with each deafening crack of thunder." This augmented description introduces highly specific, synchronized temporal events (e.g., "the ground shuddering beneath her feet with each deafening crack of thunder") that the corresponding vibration signal cannot guarantee to accurately reflect. This limitation indicates that the challenge lies not only in data scarcity but also in the lack of accurately aligned text-signal pairs. Motivated by this observation, our work emphasizes filtering and rewriting textual descriptions to better reflect vibration-oriented characteristics.
``A thunderstorm is happening.''
\subsection{Generative Paradigms for Physical Signal Generation}

Early approaches to physical signal generation were primarily based on variational autoencoders (VAEs) \cite{girin2022dynamical} and generative adversarial networks (GANs) \cite{brophy2023generative}. These models provided initial solutions for signal reconstruction and synthesis. However, their ability to model complex cross-modal sequences remains limited. In particular, when long-range temporal dependencies and fine-grained semantic control are required, these methods struggle to maintain both generation quality and semantic consistency \cite{bond2021deep, dhariwal2020jukebox, borsos2023audiolm, wen2022transformers}.

With the rapid development of text-conditioned generation, more effective paradigms have emerged for signal synthesis. Take text-to-audio generation\cite{zhao2025ai} for example. Two representative routes
have emerged. The first is based on autoregressive (AR) modeling over discrete tokens, as exemplified by AudioGen \cite{kreuk2022audiogen}. In this framework, waveform signals are first quantized into discrete tokens, which are then generated sequentially by a Transformer decoder conditioned on text. The second approach is diffusion-based, such as AudioLDM \cite{liu2023audioldm}, where generation is performed in a continuous latent space learned by a VAE, and guided by text through a shared language–audio representation. These two paradigms form the foundation of modern text-to-audio generation.

In the haptic domain, HapticGen \cite{sung2025hapticgen} introduced the first generative framework for text-to-vibration synthesis. It adopts an autoregressive Transformer adapted from the MusicGen \cite{copet2023simple} and AudioGen \cite{kreuk2022audiogen} family, and trains it on a haptic dataset bootstrapped from WavCaps. This work demonstrates the feasibility of text-to-vibration generation. However, the limitations of AR models become more pronounced for long-duration signals, such as 10-second sequences. Since inference is strictly causal and performed token by token, the model lacks a global view of the sequence. This leads to error accumulation and weak long-range structural coherence \cite{devlin2019bert, gu2017non, borsos2023soundstorm}. In addition, the global temporal envelope is difficult to model, often resulting in repetitive or over-smoothed patterns \cite{liu2023audioldm}.

Neuroscientific studies \cite{bensmaia2003vibrations, manfredi2014natural} suggest that perception of mid-to-high-frequency(40-400 Hz) vibrations is primarily governed by temporal dynamics, including rhythmic patterns and envelope variations, rather than individual sample points. Therefore, accurately modeling temporal structure and global envelope under text conditions is critical for text-to-vibration generation. To address this challenge, we propose HapticLDM, the first latent diffusion model for vibration generation. Unlike autoregressive methods, our approach performs global denoising in latent space, enabling modeling under a broader receptive field. By adapting the Stable Audio framework \cite{evans2025stable} for 10-second vibration synthesis, HapticLDM generates signals with more coherent temporal dynamics and improved alignment with textual descriptions.

%% file: 3_method.tex
\section{Methodology}

\subsection{Semantic Filtering for Haptic-Effective Text Inputs}
\label{sec:III-A}
To construct effective text–vibration training pairs, we first analyze the semantic composition of textual descriptions. Prior work in natural language processing shows that text contains heterogeneous semantic components, including action-related structures (e.g., predicates and events) \cite{palmer2005propbank}, affective expressions \cite{liu2012sentiment}, and abstract semantic content represented in lexical spaces \cite{mikolov2013word2vec}. Affective states can further be characterized along an arousal dimension, distinguishing high-arousal from low-arousal expressions \cite{russell1980circumplex}. Based on these observations, we categorize textual inputs into four types: action-related descriptions, high-arousal emotions, low-arousal emotions, and abstract concepts.

Haptic interfaces are particularly effective for conveying action-related information and salient affective cues. Action-related events can be encoded into vibrotactile signals through temporal patterns and force variations \cite{srinivasan1997haptics,choi2013vibrotactile}. Typical examples include patterns such as “rhythmic pulsing,” “engine starting,” and “alarm vibration.” In addition, variations in intensity and rhythm can reliably convey high-arousal emotions in affective haptics \cite{eid2015affective,seifi2013first}, such as “shock” or “trembling.”

Consistent with the results in Section \ref{sec:exp1-result}, we find that action-related descriptions and high-arousal emotions are more reliably represented and recognized in vibration signals. In contrast, low-arousal emotions and abstract concepts are difficult to encode using vibrotactile feedback. Therefore, we retain action-related descriptions and high-arousal emotions as effective inputs for text-to-vibration generation.

\subsection{Dataset Construction}
\input{tables/table1}
\input{tables/table2}

Dataset construction is aligned with a three-stage model training paradigm: pretraining, supervised fine-tuning (SFT), and direct preference optimization (DPO). Initially, a large-scale pretraining dataset is assembled to establish general text-vibration correspondence, where inherently invalid speech-related samples are removed. Subsequently, tactile-oriented filtering and objective rewriting are applied to source data to construct an SFT dataset, thereby achieving modal alignment. Finally, a preference dataset sourced from HapticGen \cite{sung2025hapticgen} is adopted for DPO training to further align generated signals with human perception. As preference data is directly utilized without further modification, detailed construction procedures outlined below focus on pretraining and SFT datasets.

\subsubsection{Pretraining DataSet: Audio-to-Haptic Conversion}

A large-scale pretraining dataset is first constructed from WavCaps\cite{mei2024wavcaps}. To begin with, speech-related samples are removed by Qwen2.5-32B\cite{qwen2.5}. In WavCaps, a considerable portion of the data describes speech content, such as speaking, talking, or monologues. However, speech is difficult to be reproduced effectively through vibrotactile feedback. And training on such samples may introduce noise, which is useless for meaningful vibration genration. Therefore, text-audio pairs whose captions contain speech-related content are filtered out. After this step, the dataset size is reduced from 400K to 343K.

Remaining audio data are then converted into vibration signals through audio-to-haptic conversion. Following prior work\cite{kim2023sound,sung2025hapticgen}, amplitude envelope is extracted from each audio waveform and converted into a one-dimensional vibration signal, which is used as a reference haptic signal during pretraining. During this conversion process, the sampling rate is set to 8 kHz, consistent with the PCM output specification of Meta Haptics Studio\cite{meta_haptics_studio}. The converted signals can be directly deployed and evaluated on Quest 3 devices. This configuration reduces sequence length and inference cost while remaining sufficient to capture the perceptually relevant components of low-frequency vibrotactile signals. In addition, each haptic sample is fixed to 10s to remain consistent with HapticGen, enabling direct comparison in later experiments. The paired text is adopted directly from the original audio caption. Different from HapticGen, no caption augmentation is applied at this stage. The original audio captions are retained to preserve large-scale event-level corresponding relations, meanwhile avoiding additional linguistic noise introduced by LLM rewriting\cite{krishna2023paraphrasing}. As a result, the pretraining dataset provides broad but relatively clean supervision for learning general text-vibration corresponding relations.

\subsubsection{SFT DataSet: Tactile-oriented Filtering and Objective Rewriting}
The SFT dataset is constructed from two sources: the pretraining dataset and HapticCap \cite{hu2025hapticcap}. The former provides large-scale event descriptions, while the latter provides human-written descriptions of vibration signals. In HapticCap, each vibration is described from three perspectives: sensory, emotional, and associative. Although high-arousal emotional text can be valid in principle, such text is not available in either dataset for SFT. In particular, the emotional annotations in HapticCap mainly describe users’ subjective affective responses after experiencing the vibration, rather than the signal’s observable dynamic properties. Therefore, they are not used in SFT and are removed from the dataset. In this work, supervision is restricted to text with clear tactile orientation.

For the pretraining dataset, dynamic orientation filtering is applied with Qwen2.5-32B. Captions that do not contain explicit physical processes or clear temporal dynamics are removed. In particular, abstract descriptions and emotion-related descriptions are excluded, since they provide limited guidance for vibration generation in the current data setting. This step retains only text--vibration pairs with clearer action progression or temporal change.

For HapticCap, only sensory and associative descriptions are retained. These descriptions are then rewritten with Qwen2.5-32B into objective, vibration-oriented text. Subjective rhetoric and personal judgments are suppressed. At the same time, dynamic properties of the vibration signal, such as onset, intensity growth, repetition, and decay, are made more explicit. Concrete dynamic events, such as alarm-clock vibration, are also expressed more clearly.

According to the size of processed HapticCap subset, an equal number of samples are then selected from filtered pretraining dataset randomly. These two processed datasets are combined in a 1:1 ratio to form the SFT dataset, consisting of a total of 114k samples. In this way, the SFT dataset balances the coverage of large-scale event data and the precision of human-written haptic descriptions, while providing clearer dynamic supervision for fine-tuning. The effects of tactile-oriented filtering and objective rewriting are illustrated in Table \ref{tab:physical_filtering_examples} and Table \ref{tab:objective_rewriting_examples}. 

The system prompts used in these two processes are provided in Appendix \ref{sec:appendix}.

\subsection{Model Architecture}
To satisfy the requirement of dynamic consistency, HapticLDM is built on the overall architecture of Stable Audio \cite{evans2025stable}, with its latent diffusion pipeline adapted for text-to-vibration generation. The model consists of two main components: latent-space modeling with a variational autoencoder (VAE), and vibration generation in latent space with a Diffusion Transformer (DiT). Text conditioning is introduced through a T5-base encoder \cite{raffel2020exploring}.

To compress high-dimensional waveforms, a VAE based on the Oobleck backbone \cite{evans2025stable} is employed. This choice follows the waveform autoencoding design of Stable Audio\cite{evans2025stable} and is suitable for one-dimensional vibration signals. Its hierarchical encoder-decoder architecture enables substantial temporal compression while preserving information needed for waveform reconstruction. The model is initialized with pretrained weights \cite{evans2025stable} and then adapted to vibration data. The input consists of normalized one-dimensional vibration waveforms sampled at 8 kHz and fixed to a duration of 10 seconds. Through this module, each waveform is mapped into a latent space with a downsampling ratio of 2048 and 64 latent channels, which substantially reduces sequence length. To improve reconstruction quality, a composite loss is adopted. The overall VAE objective, denoted by $\mathcal{L}_{\mathrm{total}}$, is defined as the sum of four components:

\begin{equation}
\mathcal{L}_{total} = \mathcal{L}_{L1} + \mathcal{L}_{STFT} + \mathcal{L}_{adv} + \mathcal{L}_{KL}.
\end{equation}

The waveform $\mathrm{L}_{1}$ loss, denoted by $\mathcal{L}_{\mathrm{L1}}$, measures the absolute difference between the original vibration signal $x$ and the reconstructed signal $\hat{x}$ in the time domain:

\begin{equation}
\mathcal{L}_{L1} = \mathbb{E}[|x - \hat{x}|].
\end{equation}

To preserve frequency-domain fidelity, a multi-resolution short-time Fourier transform (STFT) loss $\mathcal{L}_{\mathrm{STFT}}$ is further introduced \cite{yamamoto2020parallel}. This loss is computed over $M$ analysis windows so that spectral differences can be penalized at multiple resolutions. For each resolution $m$, spectral convergence and log-magnitude distance are combined as follows:

\begin{equation}
\begin{split}
\mathcal{L}_{STFT} &= \frac{1}{M} \sum_{m=1}^{M} \Biggl( \frac{\| |\mathrm{STFT}_{m}(x)| - |\mathrm{STFT}_{m}(\hat{x})| \|_{F}}{\| |\mathrm{STFT}_{m}(x)| \|_{F}} \\
&\quad + \| \log|\mathrm{STFT}_{m}(x)| - \log|\mathrm{STFT}_{m}(\hat{x})| \|_{1} \Biggr).
\end{split}
\end{equation}

Perceptual realism is further encouraged by introducing an adversarial loss $\mathcal{L}_{\mathrm{adv}}$ \cite{kong2020hifi}. Under this objective, reconstructed signals are penalized when they are identified as artificial by the discriminator $D$:

\begin{equation}
\mathcal{L}_{adv} = \mathbb{E}[(1 - D(\hat{x}))^{2}].
\end{equation}

A separate objective is used to train the discriminator $D$. In this step, both real vibration signals $x$ and reconstructed signals $\hat{x}$ are evaluated, and the discriminator loss $\mathcal{L}_{\mathrm{dis}}$ is defined as:

\begin{equation}
\mathcal{L}_{dis} = \mathbb{E}[(D(x) - 1)^{2} + D(\hat{x})^{2}].
\end{equation}

During optimization, the VAE parameters are updated with $\mathcal{L}_{\mathrm{total}}$, while the discriminator parameters are updated with $\mathcal{L}_{\mathrm{dis}}$ in an alternating manner. This training scheme helps maintain stable convergence.

The encoded latent distribution $q(z|x)$ is further regularized toward the standard normal prior $p(z)$ through a Kullback--Leibler (KL) divergence loss $\mathcal{L}_{\mathrm{KL}}$ \cite{kingma2013auto}. This term encourages a continuous latent space for subsequent diffusion modeling:

\begin{equation}
\mathcal{L}_{KL} = \mathbb{E}[D_{KL}(q(z|x) \parallel p(z))].
\end{equation}

Empirically, these combined objectives preserve temporal envelope structure in the adapted latent representation, which is important for haptic perception.

\input{figs/fig1}
Once the VAE has learned a continuous latent space, a DiT-based diffusion model is trained on the compressed latent representations. Text conditioning is introduced through cross-attention, using embeddings from a T5-base encoder \cite{raffel2020exploring}. Diffusion is performed directly in latent space. Compared with direct waveform generation, latent-space modeling reduces temporal resolution and simplifies the denoising problem. As a result, the model can focus more on slowly varying dynamic structure, such as rhythm and envelope change, instead of sample-level fluctuations. This is well suited to vibration generation, where perceptual meaning is conveyed primarily through temporal evolution rather than fine waveform detail. Following Stable Audio \cite{evans2025stable}, the diffusion model is trained with v-parameterization \cite{salimans2022progressive}, implemented as a mean squared error loss between the target velocity $v$ and the predicted velocity $v_{\theta}$:

\begin{equation}
\mathcal{L}_{diff} = \mathbb{E}_{z_0, \epsilon, t} \left[ \| v - v_{\theta}(z_t, t, c) \|_{2}^{2} \right],
\end{equation}
where $z_t$ denotes the noisy latent at timestep $t$, $c$ denotes the text condition, and the target velocity is defined according to continuous-time schedules. After sampling, the generated latents are decoded into waveform space by the VAE decoder. A post-processing amplitude validity check is then applied to discard overly weak outputs. Specifically, the absolute waveform values are evaluated, and both the maximum amplitude and the 99.5th percentile amplitude are computed. A generated signal is retained only when both values exceed predefined minimum thresholds; otherwise, it is discarded.

\subsection{Training Strategy}

The training process of HapticLDM is illustrated in Fig.\ref{Fig.1}. The training pipeline consists of four stages: VAE adaptation, diffusion pretraining, supervised fine-tuning, and preference optimization\cite{wallace2024diffusion}. All experiments are conducted on 4 NVIDIA RTX 5880 Ada GPUs. For all datasets, the data are split into training and validation sets with a ratio of 9:1.

Initially, VAE adaptation is executed on a comprehensive collection of vibration signals, which comprises all data from both pretraining and SFT datasets. During this process, batch size of 20 is utilized. Since VAE weights are derived from audio-based pretraining, inherent mismatch may exist between original audio-oriented latent space and target vibration signals. Consequently, weights are tuned on vibration data so that compressed representations can better preserve vibration-specific temporal structures, especially envelope patterns that are considered important for haptic perception.

After VAE adaptation, the diffusion model is trained in three stages. In the first stage, pretraining is performed on the pretraining dataset with a batch size of 512. This stage is used to learn general text-vibration correspondence from large-scale paired data. Although the converted vibration signals and original captions are not as precise as human-written haptic descriptions, they still provide useful supervision on diverse physical events and temporal patterns. In this way, the model is first given general coverage before later refinement.

In the second stage, the pretrained diffusion model is further fine-tuned on the SFT dataset with a batch size of 512. Compared with the pretraining dataset, the SFT dataset provides cleaner and more vibration-oriented supervision after tactile-oriented filtering and objective rewriting. This stage is used to refine the alignment between textual dynamics and generated vibration patterns. In particular, clearer physical processes and temporal cues are emphasized, and vague or weakly dynamic descriptions are reduced. As a result, the signals generated by proposed method are expected to be more consistent with the dynamic content of the prompts.

In the third stage, preference optimization is performed on the DPO dataset constructed from HapticGen\cite{sung2025hapticgen}, with a batch size of 128. This stage is introduced after supervised fine-tuning to further improve generation quality at the preference level. Although previous stages mainly rely on paired supervision, they do not directly model relative preference between candidate outputs. DPO training is used to address this issue. By learning from pairwise preference signals, the model is further guided toward outputs that are better aligned with human preference in terms of vibration quality and text correspondence.

Overall, the training strategy is designed in a progressive manner. The VAE is first adapted to establish a vibration-oriented latent space. The diffusion model is then pretrained on large-scale data, refined with higher-quality supervision, and finally aligned with preference signals. This staged design helps combine data scale, text quality, and preference alignment within a unified training pipeline.

%% file: tables/table1.tex
\begin{table}[t]
\centering
\caption{Examples of Tactile-Oriented Filtering.}
\label{tab:physical_filtering_examples}
\renewcommand{\arraystretch}{2}
\setlength{\tabcolsep}{4pt}
\begin{tabular}{p{0.42\linewidth} p{0.48\linewidth}}
\toprule
\textbf{Input Text} & \textbf{Reason for Filtering Out} \\
\midrule
A bicycle is producing a sound. 
& `A bicycle is producing a sound' is too vague and does not specify a distinct physical action or texture. \\

String ensemble is playing a sequence of major chords. 
& `String ensemble playing chords' describes a melodic musical performance which loses its meaning in RMS envelope extraction. \\

Noise is being removed from a sound file. 
& `Noise removal' is a process description without a clear sound/vibration source. \\
\bottomrule
\end{tabular}
\end{table}

%% file: tables/table2.tex
\begin{table}[t]
\centering
\caption{Examples of Objective Rewriting.}
\label{tab:objective_rewriting_examples}
\renewcommand{\arraystretch}{2}
\setlength{\tabcolsep}{4pt}
\begin{tabular}{p{0.48\linewidth} p{0.42\linewidth}}
\toprule
\textbf{Input Text} & \textbf{Rewritten Text} \\
\midrule
This reminds me when I am waking up early in the morning with my alarm and switch it off 
& Waking up early in the morning with an alarm and switching it off \\

It is like continues low frequency vibration with different pitch levels 
& Continuous low frequency vibration with different pitch levels \\

I would descibe the sensation as stimulative. 
& Stimulative sensation \\
\bottomrule
\end{tabular}
\end{table}

%% file: figs/fig1.tex
\begin{figure*}[ht] 
    \centering
    \includegraphics[width=\linewidth]{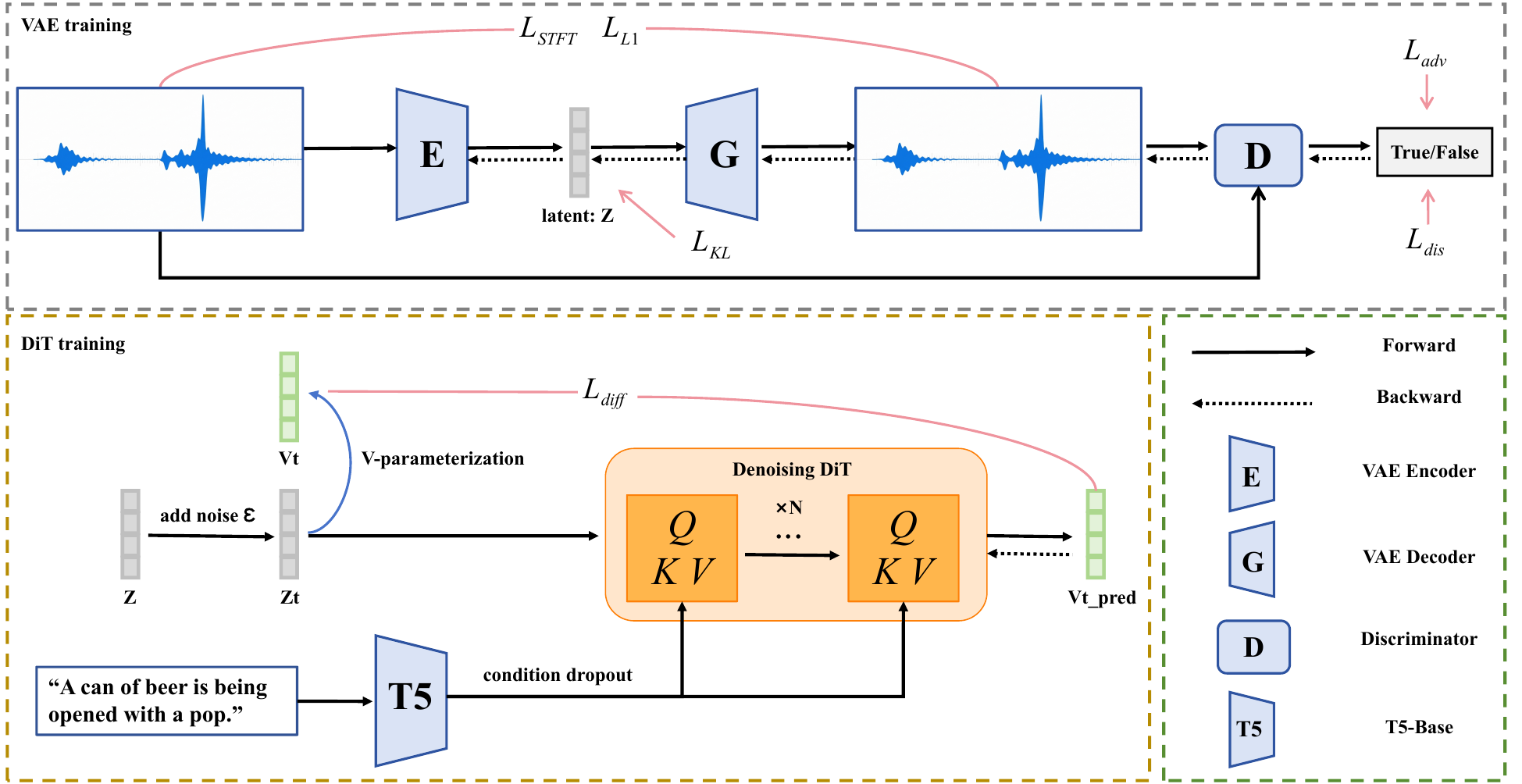}
    \caption{The overall training framework of HapticLDM. Top (VAE Training): The VAE (comprising Encoder E and Decoder G) and the Discriminator (D) are trained alternately in an adversarial manner. Specifically, D is updated to distinguish real vibrations from reconstructed ones using $\mathcal{L}_{dis}$, while the VAE is optimized to fool D via $\mathcal{L}_{adv}$, alongside the reconstruction losses ($\mathcal{L}_{STFT}, \mathcal{L}_{L1}$) and the KL divergence ($\mathcal{L}_{KL}$). Bottom (DiT Training): In the latent space learned by the VAE, a Diffusion Transformer (DiT) is trained to predict the velocity target $V_{t}$ guided by text conditions. The text embedding acts as Keys (K) and Values (V) via cross-attention. Condition dropout is randomly applied to the text embeddings during this training phase. The diffusion model is optimized using the $v$-parameterization loss ($\mathcal{L}_{diff}$).}
    \label{Fig.1}
\end{figure*}

%% file: 4_experiments.tex
\input{figs/exp_process}

\section{Experiment}

In this section, we design experiments to evaluate both the dataset and the proposed method for text-to-vibration generation. The experimental setup is described as follows.

\subsection{Verification of High-Quality Data Pairs}
\label{sec:exp-1}
As described in Section \ref{sec:III-A}, the dataset is divided into four semantic categories: action-related descriptions, high-arousal emotions, abstract concepts, and low-arousal emotions. To evaluate whether vibration signals can accurately represent textual semantics, we randomly select five samples from each category in both the HapticCap and HapticGen datasets, resulting in 20 groups in total.

For each text, its corresponding vibration signal is treated as a positive sample. A vibration signal associated with another semantically similar text from the same category is used as a negative sample. This process yields a test set of 40 samples (20 positive and 20 negative).

Twenty participants rate the matching degree between text and vibration on a 5-point Likert scale. A score above 3 indicates that the vibration effectively conveys the semantic content. The difference in scores between positive and negative samples reflects discriminability. Samples with both high matching scores and clear discriminative differences are considered high-quality.

\subsection{Verification of the Superiority of Proposed Method}
\label{sec:exp-2}
As shown in Fig. \ref{fig:exp_process}, we compare our method with HapticGen using the same prompt themes: sport, interaction, game, and simulation. A total of 25 prompts are randomly generated by GPT-5.4 \cite{openai2026gpt54}. Corresponding vibration signals are then generated by both the proposed method and HapticGen. To evaluate efficiency and randomness, each method generates five vibration signals per prompt. An A/B test is then conducted as follows:

\begin{enumerate}
\item The 25 prompts are divided into three groups.
\item In each group, methods A and B are randomly assigned to HapticLDM or HapticGen, and participants are blinded to the assignment.
\item For each prompt, five samples from method A and five from method B are presented. Participants evaluate whether each vibration is suitable for the prompt and vote accordingly.
\item Participants rate both methods within each group on three aspects—satisfaction, expressiveness, and realism—using a 5-point Likert scale. They also indicate an overall preference.
\item After evaluation, a semi-structured interview is conducted to collect feedback on user experience, impressions, and suggestions.
\end{enumerate}

A total of 30 participants from diverse professional backgrounds, age groups, and regions are recruited. Among them, 10 are experts in haptic feedback or human–computer interaction.

%% file: figs/exp_process.tex
\begin{figure*}[ht] 
    \centering
    \includegraphics[width=\linewidth]{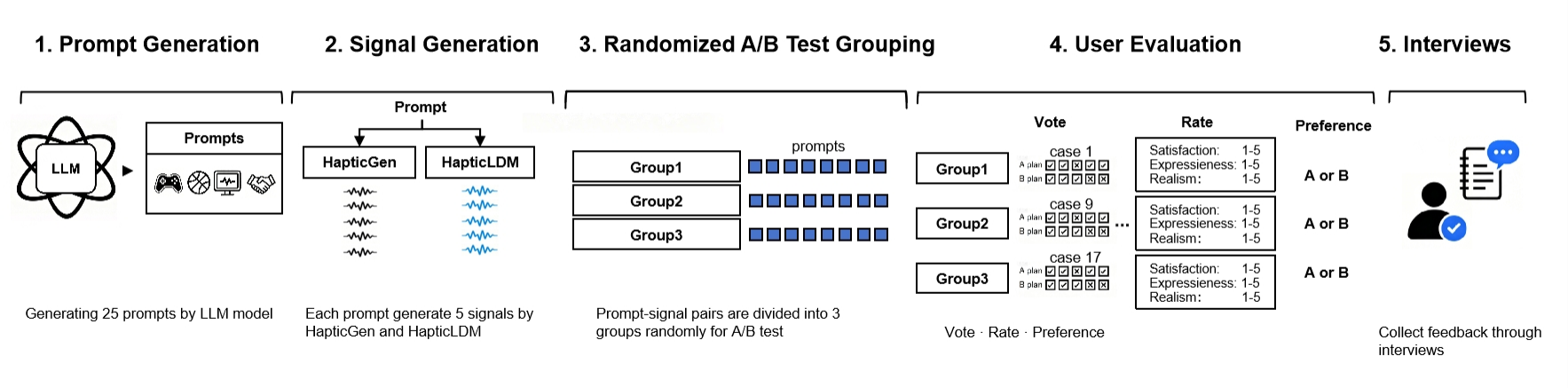}
    \caption{The workflow of random A/B test for performance comparison between HapticLDM and HapticGen
    }
    \label{fig:exp_process}
\end{figure*}

%% file: 5_results.tex
\section{Results}

\subsection{The quality of data with different prompt type}

 As introduced in Section \ref{sec:exp-1}, data pairs from four categories—action-related descriptions, high-arousal emotions, abstract concepts, and low-arousal emotions—are evaluated in terms of suitability. Scores are reported as mean ± standard deviation. According to Fig.~\ref{Fig.3} and Table~\ref{tab:exp2table}, action-related descriptions and high-arousal emotions outperform abstract concepts and low-arousal emotions. Specifically, action-related data achieve a score of $3.44 \pm 1.21$, while high-arousal emotions reach $3.23 \pm 1.22$. In contrast, abstract concepts and low-arousal emotions perform worse, with scores of $2.76 \pm 1.22$ and $2.68 \pm 1.31$, respectively
 \input{tables/exp2table}
 \label{sec:exp1-result}
Fig.~\ref{Fig.3} also illustrates recognizability by comparing positive and negative samples. Action-related descriptions show the largest score difference between positive and negative samples ($1.04$), indicating strong discriminability. In contrast, although high-arousal emotions achieve a relatively high score ($3.23$), their negative samples also receive high ratings ($2.98$), suggesting limited distinguishability. Overall, action-related data exhibit both high suitability and strong discriminability, and are therefore considered high-quality data for model training. In contrast, abstract concepts and low-arousal emotions lack explicit dynamic mappings and are less suitable for text-to-vibration generation.
\input{figs/fig3}

\subsection{The performance comparison between HapticGen and HapticLDM}
The results of Experiment \ref{sec:exp-2} are summarized in Fig.~\ref{Fig.4} and Table~\ref{tab:exp3table}. Quantitative analysis of voting results and Likert-scale ratings shows a consistent preference for HapticLDM over HapticGen across all test groups. As illustrated in Fig.~\ref{Fig.4}(a), HapticLDM outperforms HapticGen in all aspects, including satisfaction, expressiveness, and realism. The largest improvement is observed in expressiveness, where HapticLDM achieves a score of $3.86$ compared to $2.54$ for HapticGen. Interview feedback further supports this result. Many participants report that vibrations generated by HapticGen often have insufficient intensity and do not match the input text well. Statistical analysis using t-tests (Table~\ref{tab:exp3table}) confirms that HapticLDM significantly outperforms HapticGen in all three aspects (p \textless 0.001). The overall score of $3.79$ demonstrates the effectiveness of HapticLDM for text-to-vibration generation. To account for the stochastic nature of generative models, adoption rates are reported in Fig.~\ref{Fig.4}(b). HapticLDM achieves an average adoption rate above 70\%, significantly higher than the 40.5\% achieved by HapticGen.

\input{tables/exp3table}

\input{figs/fig4}

Interview results also highlight improvements in rhythm and frequency matching, with HapticLDM providing more accurate feedback patterns. However, several limitations are noted. Experts point out the lack of precise “aftershock” (damping) effects across different materials, as well as insufficient low-frequency performance for simulating “swaying” or “heavy shaking” in complex interactions. Hardware constraints, such as actuator performance, are also identified as a limiting factor. From a usability perspective, participants generally agree that the generative system is valuable for rapid prototyping and can serve as an “inspiration aid” for haptic design. However, they note that the “hit rate”—the frequency of generating satisfactory outputs—still requires improvement to meet professional standards.

%% file: tables/exp2table.tex
\begin{table}[h]
\centering
\caption{The statistic results of evaluations on positive and negative samples by categories}
\label{tab:exp2table}
\begin{tabular}{lccc}
\toprule
\textbf{Category} & \textbf{\begin{tabular}[c]{@{}c@{}}Positive samples\\ Mean$\pm$SD\end{tabular}} & \textbf{\begin{tabular}[c]{@{}c@{}}Negative samples\\ Mean$\pm$SD\end{tabular}} & \textbf{$|$Diff$|$} \\ 
\midrule
\textbf{\begin{tabular}[l]{@{}l@{}}Action-related\\ Description\end{tabular}} & \textbf{3.44$\pm$1.21} & 2.40$\pm$1.24 & \textbf{1.04} \\
High Arousal & 3.23$\pm$1.22 & 2.98$\pm$1.23 & 0.25 \\
Abstract Concepts & 2.76$\pm$1.22 & 2.12$\pm$1.09 & 0.64 \\
Low Arousal & 2.68$\pm$1.31 & 2.33$\pm$1.37 & 0.35 \\ 
\bottomrule
\end{tabular}
\end{table}

%% file: figs/fig3.tex
\begin{figure}[ht] 
    \centering

    \includegraphics[width=\linewidth]{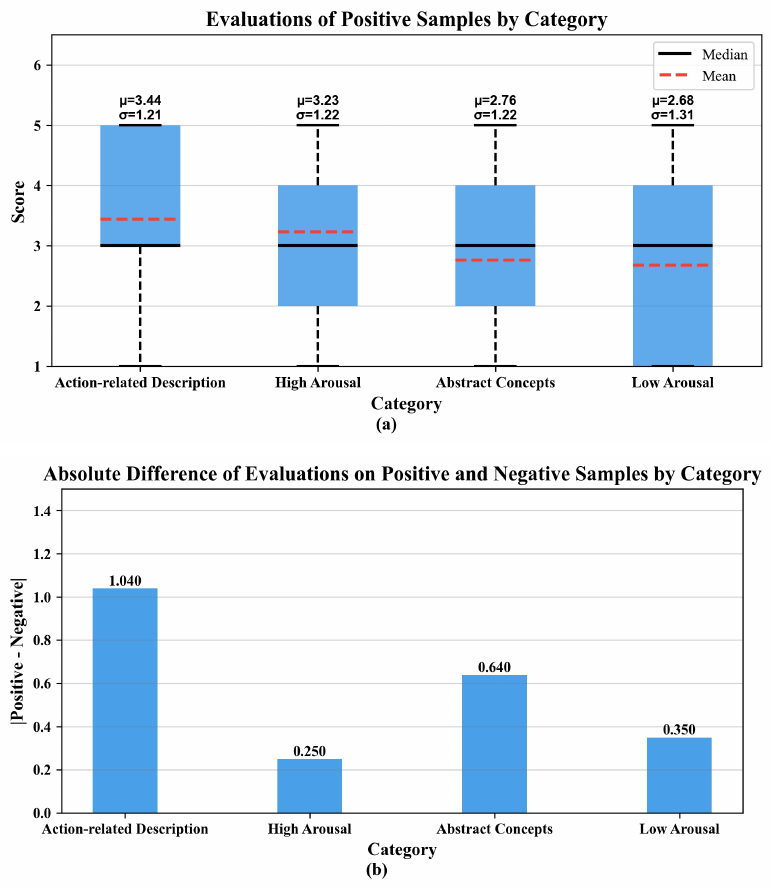}

    \caption{(a) Box plot of the 5-Likert scale scoring results for 4 types data pairs as action-related descriptions, high-arousal emotions, abstract concepts, and low-arousal emotions. (b) The histogram of absolute difference between positive and negative samples on 5-Likert scale scores.} 

    \label{Fig.3}
\end{figure}

%% file: tables/exp3table.tex
\begin{table}[h]
\centering
\caption{T test results of evaluations on HapticLDM and HapticGen}
\label{tab:exp3table}
\begin{tabular}{lcccc}
\toprule
\textbf{Metric} & \textbf{Student's t} & \textbf{Cohen's d} & \textbf{p} & \textbf{Significant} \\ 
\midrule
Satisfaction & 7.834 & 1.713 & $< .001$ & $***$ \\
Expressiveness & 8.477 & 1.880 & $< .001$ & $***$ \\
Realism & 7.284 & 1.559 & $< .001$ & $***$ \\ 
\bottomrule
\multicolumn{5}{l}{\footnotesize $^{***} p < 0.001$, $^{**} p < 0.01$, $^{*} p < 0.05$} \\ 
\end{tabular}
\end{table}

%% file: figs/fig4.tex
\begin{figure}[ht] 
    \centering

    \includegraphics[width=\linewidth]{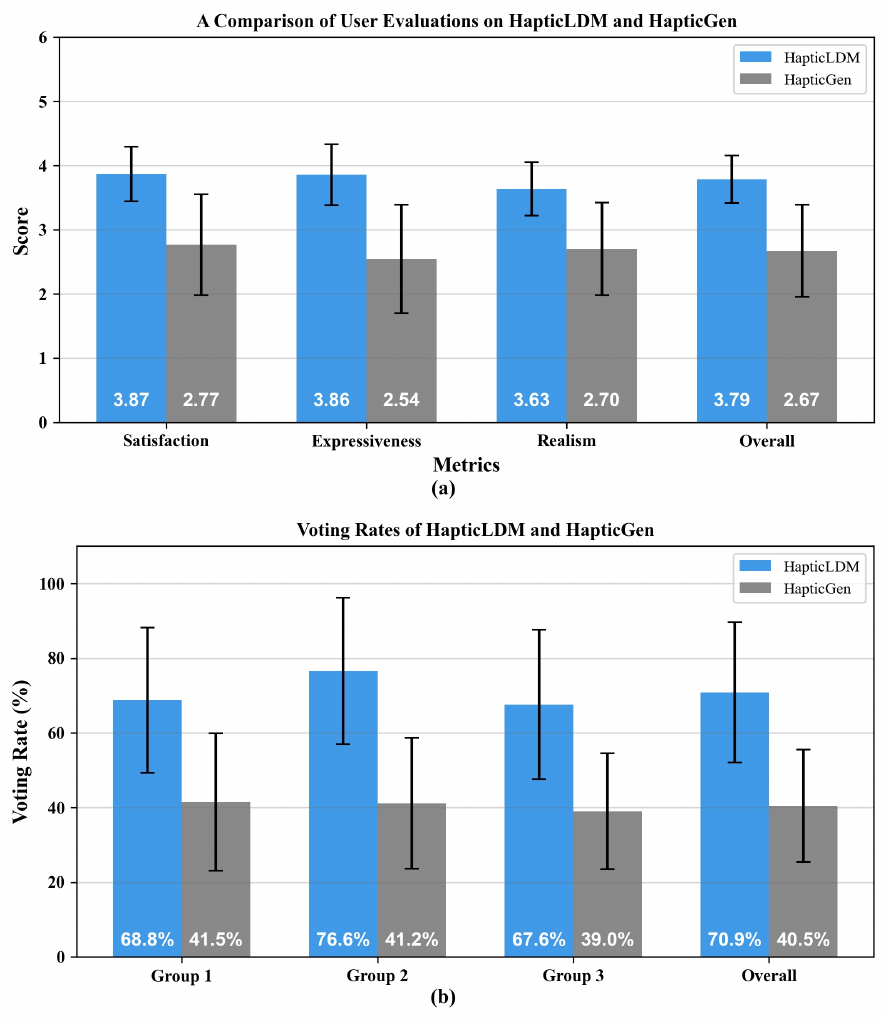}

    \caption{Results of the Experiment 4.3, comparing the performance of Haptic LDM and Haptic Gen by A/B test. (a) The evaluation of methods on different aspects as satisfaction, expressiveness and realism. (b) The adopting rate of methods in different groups.} 

    \label{Fig.4}
\end{figure}

%% file: 6_disscusion.tex
\section{Discussion}
Further reflections on the proposed method, including its main contributions and limitations, are discussed as follows.

\subsection{Reflections on Contributions}

This study examines two key aspects: \textit{data validity} and the \textit{generative method}.

First, data validity is identified as a critical factor in constructing high-quality datasets. Compared with abstract concepts and low-arousal emotions, action-related descriptions and high-arousal emotions exhibit clearer perceptual mappings. These categories achieve better performance in vibration generation and are therefore more suitable for model training. Experimental results further confirm the difficulty of evaluating data associated with abstract concepts and low-arousal emotions. Participants report that it is challenging to associate such descriptions with concrete vibration patterns. For example, prompts such as “Just finishing an exam after doing no study” are difficult to translate into meaningful tactile feedback.

In addition, prompt specificity significantly affects evaluation outcomes. Participants tend to assign higher satisfaction scores to coarse or underspecified prompts (e.g., “a rhythmic heartbeat”), as multiple vibration patterns can reasonably match the same description. For instance, both slow and fast rhythms may be considered appropriate. In contrast, detailed prompts require precise control over rhythm, texture, and intensity, leading to stricter evaluation criteria.

Second, we propose HapticLDM with a progressive alignment training strategy for high-fidelity and long-duration vibration generation. Evaluations involving 30 participants demonstrate that the proposed method outperforms baseline approaches in text-to-vibration tasks, achieving higher satisfaction, realism, and expressiveness. These gains come from the diffusion-based formulation. It refines signals in continuous space and preserves global temporal structure, resulting in more stable long-duration generation than auto-regressive models, which often accumulate errors. The iterative process also enables fine-grained conditioning on text. In addition, sampling from a noise prior captures the multi-modal nature of vibration signals and avoids over-smoothed outputs. The progressive alignment strategy further improves the consistency between text and vibration. However, the iterative sampling process increases computational cost and limits real-time use.

\subsection{Limitations and Future work}
\textbf{The shortcomings in the quality of generation. }
As noted by participants, the current generative method still faces challenges in accurately simulating material-specific textures and complex physical feedback. The precision of rhythmic patterns and frequency representation also requires improvement. One key limitation lies in the lack of structured and detailed textual descriptions, which constrains the model’s ability to capture fine-grained temporal characteristics. From a modeling perspective, both the network architecture and training strategy require further refinement to better control rhythm and frequency features, thereby improving generation quality. In addition, hardware plays a critical role in vibrotactile feedback, as it fundamentally determines the upper performance bound of the system. Factors such as the number and type of actuators, their spatial arrangement, and supporting structures all influence the resulting feedback. Although the proposed method considers the frequency response of the device, it remains insufficient for generating complex physical interactions and material-specific textures. Incorporating hardware characteristics into the generative process is therefore an important direction for future work in physical signal generation.

\textbf{Multi-modal scene matching. }
Haptic feedback generation from multimodal data remains a challenging and important research problem. Compared with audio and visual modalities, text exhibits a larger representational gap with vibration signals, making it more difficult to establish reliable sensory mappings between text and haptics. In this study, several participants noted that visual cues, such as images or videos, could assist in evaluating vibration effects. However, incorporating multimodal inputs introduces additional complexity, as it requires precise temporal alignment across different modalities.

\textbf{Complex scene and unfixed-length sequence generation. }
Experimental results indicate that the proposed method struggles to reconstruct complex vibration patterns involving multiple interacting factors, such as “Rowing a boat on choppy water with strong waves.” The temporal dynamics in the generated signals are often insufficiently pronounced or inaccurate. These limitations are primarily attributed to both the dataset and the training strategy. In particular, only a small portion of prompts explicitly describe detailed temporal dynamics, which restricts the model’s ability to learn fine-grained variations. Moreover, extracting structured feature representations from text across different dimensions remains challenging, limiting the design of effective control and training strategies. In addition, several participants noted that fixing the generation length to 10 seconds is suboptimal. In some cases, the generated signals appear repetitive due to prolonged patterns, while in others, the temporal evolution is incomplete within the fixed duration. These observations suggest that adaptive length generation conditioned on prompt semantics is necessary for more realistic haptic synthesis.

\textbf{Adopting rate. }
Efficiency is a critical factor for the practical adoption of generative methods. Although the proposed approach can produce high-quality vibration signals, the experimental adoption rate of 50\% indicates a non-negligible failure rate. Due to the inherent stochasticity of the generation process, the quality of the outputs remains inconsistent. Therefore, improving reliability and ensuring stable generation quality remain important challenges.

\textbf{Prompt control. }
Analysis of failure cases indicates that the proposed method is insensitive to fine-grained feature prompts, such as “3/4 beat” and “rapid decrease in intensity,” and thus fails to generate precise vibration patterns corresponding to these specifications. This limitation primarily arises from the current generation strategy. Specifically, the model processes the entire prompt holistically during generation, without explicit control over individual features. As a result, it favors global reconstruction and often loses fine-grained temporal characteristics. Incorporating feature-level control mechanisms could further improve the precision and controllability of the generated results.

%% file: 7_Conclusion.tex
\section{Conclusion}
In this study, we introduce HapticLDM to enable more efficient and immersive haptic feedback generation. The improvements are achieved from both data and modeling perspectives.

From the data perspective, experimental results show that action-related descriptions provide more reliable representations for vibration synthesis. By filtering and rewriting textual prompts, vibration-oriented characteristics are enhanced. The processed data exhibits more precise dynamic characteristics and perceptual mappings, which guaranties more effective model training. From the modeling perspective, user evaluations demonstrate the superiority of diffusion-based models in generating coherent and expressive vibration signals.

In summary, the proposed method shows strong potential for simplifying haptic design and improving generation efficiency and quality. It also provides insights into generative modeling strategies for physical vibration signals.

%% file: 8_acknowledgement.tex
\section*{Acknowledgment}
This work was supported by Guangdong Provincial R\&D Program in Key Areas under Grant 2023B0101200011.

The authors would like to thank Dr. Guangyao Xu for helpful discussions and technical support.

%% file: 9_appendix.tex
\section{System Prompts for Tactile-oriented Filtering and Objective Rewriting}
\label{sec:appendix}
The primary system instructions used for tactile-oriented filtering are presented below.

\vspace{1em}
\hrule
\vspace{0.5em}
\noindent\textbf{System Prompt: Haptic Feedback Filtering}

\vspace{0.5em}
\noindent You are an expert in Haptic Feedback Design and Vibration Texture Analysis. Your task is to filter text descriptions of audio files to determine if they are suitable for accompanying a vibration signal generated via RMS envelope extraction.

\vspace{0.5em}
\noindent The audio is converted to vibration solely based on amplitude and rhythm. Pitch and melody are lost. The goal is to keep descriptions where the user, upon feeling the vibration and reading the text, feels high consistency.

\vspace{0.5em}
\noindent\textbf{Classification Dimensions:}

\noindent\textbf{Dimension 1 (Objective Perception):} Describes intensity, envelope, texture, or rhythm (e.g., ``pulsing'', ``banging'', ``steady hum''). $\rightarrow$ \textbf{KEEP}

\noindent\textbf{Dimension 2.1 (Concrete/Physical Association):} Describes a specific physical object, material interaction, or biological action that creates a distinct physical vibration pattern and allows humans to directly associate it with a consistent vibration scenario when reading this description.

\noindent\textbf{Dimension 2.2 (Abstract/Emotional):} Describes emotions, mood, functional meaning, or internal states (e.g., ``scary'', ``good news'', ``victory'', ``suspenseful'', ``zombie wanting brains''). $\rightarrow$ \textbf{DISCARD}

\noindent\textbf{Dimension 3 (Other):} Vague ambiance, pure silence, or visual-only descriptions without clear sound/vibration source. $\rightarrow$ \textbf{DISCARD}

\vspace{0.5em}
\noindent\textbf{Filtering Rules:}
\begin{itemize}
    \item \textbf{KEEP} if the text falls strictly into Dimension 1 or Dimension 2.1.
    \item \textbf{DISCARD} if the text relies heavily on Dimension 2.2 or Dimension 3.
    \item \textbf{DISCARD} descriptions of ``music'', ``melodies'', or ``songs'' unless they describe a rhythmic physical action (like ``drum roll''), because RMS destroys melody.
\end{itemize}

\vspace{0.5em}
\noindent\textbf{Critical Output Requirement:} \\
You must respond with only a valid json object. Do not include any other text, explanations, markdown, or formatting.\\
Your response must be EXACTLY in this format: \\
\texttt{\{``keep'': true/false, ``reason'': your explanation here\}}

\vspace{1em}
\hrule
\vspace{1em}

To further guide the model, several few-shot examples are appended to the system prompt. These examples and their expected outputs are detailed in Table \ref{tab:few_shot_examples}.

The primary system instructions used for data cleaning and objective rewriting are presented below.

\vspace{1em}
\hrule
\vspace{0.5em}
\noindent\textbf{System Prompt: Objective Rewriting and Data Cleaning}

\vspace{0.5em}
\noindent You are a strict data cleaning assistant for a haptic feedback dataset. Your task is to filter and rewrite raw user text into objective, concise captions.

\vspace{0.5em}
\noindent You must analyze the input and output a VALID JSON object.

\vspace{0.5em}
\noindent\textbf{Filtering Rules (\texttt{``keep'': true} or \texttt{false}):}
\\ Check if the input falls into one of the rejection categories below. If it does, set \texttt{``keep'': false} and provide the specific \texttt{``reason''}.

\vspace{0.5em}
\noindent\textbf{Rejection Categories:}
\begin{itemize}
    \item \textbf{``N/A'':} For placeholders like ``NA'', ``N.A.'', ``n.a'', ``n/a'', ``.'', ``None'', ``Unknown'', ``Null''.
    \item \textbf{``No Sensation'':} If it is explicitly stated that no sensation was felt (e.g., ``I don't think so'', ``No'', ``Never experienced it'', ``I did not feel it'', ``Nothing'', ``Hard to tell'').
\end{itemize}

\vspace{0.5em}
\noindent\textbf{Rewriting Standards (If \texttt{``keep'': true}):}
\\ The text must be rewritten by applying all the following standards simultaneously.

\vspace{0.5em}
\noindent\textbf{A. Data Integrity (No Hallucination - Critical):}
\begin{itemize}
    \item \textbf{NEVER} add actions, adjectives, intensities, frequencies, or speeds that are not explicitly present.
    \item Only preserve details that are already in the input text.
    \item If information is vague, it must be kept vague. Do not clarify or infer.
\end{itemize}

\vspace{0.5em}
\noindent\textbf{B. Style \& Perspective Transformation:}
\begin{itemize}
    \item \textbf{Remove conversational/reflective phrases:} Delete ``I think'', ``I feel'', ``This sensation'', ``It feels like'', ``It reminds me of''.
    \item \textbf{Remove observer perspective:} Focus on the described phenomenon itself, rather than the act of sensing.
    \item \textbf{Convert analogies into direct references:} 
    \begin{itemize}
        \item ``reminds me of EDM music'' $\rightarrow$ ``EDM music''
        \item ``feels like a heartbeat'' $\rightarrow$ ``A heartbeat''
        \item ``reminds me of my palm twitching'' $\rightarrow$ ``Palm twitching''
    \end{itemize}
    \item \textbf{Generalize personal references:} 
    \begin{itemize}
        \item ``my phone'' $\rightarrow$ ``A mobile phone''
        \item ``my hand'' $\rightarrow$ ``A hand''
    \end{itemize}
\end{itemize}

\vspace{0.5em}
\noindent\textbf{Output Format (Strict JSON):}
\begin{itemize}
    \item If keeping: \\
    \texttt{\{``keep'': true,} \\
    \texttt{``rewritten\_text'': cleaned\_string\}}
    \item If rejecting: \\
    \texttt{\{``keep'': false,} \\
    \texttt{``reason'': CATEGORY\_NAME\}} \\
    (Choose one of: ``N/A'', ``No Sensation'')
\end{itemize}

\vspace{1em}
\hrule
\vspace{1em}

To further guide the objective rewriting process, several few-shot examples are appended to the system prompt. These examples and their expected outputs are detailed in Table \ref{tab:few_shot_rewriting}.

\input{tables/Dynamic_filter}
\input{tables/Rewritten}

%% file: tables/Dynamic_filter.tex
\begin{table*}[htbp]
\centering
\caption{Few-shot examples provided for tactile-oriented filtering.}
\label{tab:few_shot_examples}
\renewcommand{\arraystretch}{1.4}
\begin{tabular}{p{0.3\textwidth} p{0.6\textwidth}}
\hline
\textbf{Input Text} & \textbf{Expected JSON Output \& Reasoning} \\
\hline
``A heartbeat sound effect, great for hospital or human sound.'' & {\ttfamily \{"keep": true, "reason": "Dimension 2.1: Heartbeat is a distinct rhythmic physical event compatible with RMS envelopes."\}} \\
\hline
``A ta da sound effect for good news or winning.'' & {\ttfamily \{"keep": false, "reason": "Dimension 2.2: 'Ta da', 'good news', and 'winning' are abstract cultural concepts/emotions, not physical textures."\}} \\
\hline
``A skillet hits someone in the head, great for cartoon or movie scene.'' & {\ttfamily \{"keep": true, "reason": "Dimension 2.1: Metal hitting head is a concrete physical impact."\}} \\
\hline
``Zombies have a thirst for brains.'' & {\ttfamily \{"keep": false, "reason": "Dimension 2.2/3: Describes an internal desire/state, not a sound-making action."\}} \\
\hline
``Strange pulsing alien noise is happening.'' & {\ttfamily \{"keep": true, "reason": "Dimension 1: 'Pulsing' describes the objective rhythm/envelope."\}} \\
\hline
``Kids are laughing.'' & {\ttfamily \{"keep": true, "reason": "Dimension 2.1: Laughter is a specific biological vocalization characterized by rhythmic bursts of amplitude, creating a distinct physical texture."\}} \\
\hline
``Someone is eating chips.'' & {\ttfamily \{"keep": true, "reason": "Dimension 2.1: 'Eating chips' implies a distinct crunching texture (sharp transients) that translates well to tactile feedback."\}} \\
\hline
``Drum kit is playing.'' & {\ttfamily \{"keep": true, "reason": "Dimension 2.1: While musical, drums are purely rhythmic and percussive, providing clear physical impacts for vibration."\}} \\
\hline
``An engine is being heard.'' & {\ttfamily \{"keep": true, "reason": "Dimension 2.1: Engines produce a continuous, rumbling physical vibration pattern."\}} \\
\hline
``Room tone and distant activity are being recorded.'' & {\ttfamily \{"keep": false, "reason": "Dimension 3: 'Room tone' and 'distant activity' are vague ambiance descriptions without clear sound/vibration source."\}} \\
\hline
\end{tabular}
\end{table*}

%% file: tables/Rewritten.tex
\begin{table*}[htbp]
\centering
\caption{Few-shot examples provided for objective rewriting and data cleaning.}
\label{tab:few_shot_rewriting}
\renewcommand{\arraystretch}{1.4}
\begin{tabular}{p{0.3\textwidth} p{0.6\textwidth}}
\hline
\textbf{Input Text} & \textbf{Expected JSON Output} \\
\hline
``The sensation is like smooth waves.'' & {\ttfamily \{"keep": true, "rewritten\_text": "Smooth waves"\}} \\
\hline
``n.a.'' & {\ttfamily \{"keep": false, "reason": "N/A"\}} \\
\hline
``This reminds me of watching a bouncy ball bounce up and down.'' & {\ttfamily \{"keep": true, "rewritten\_text": "A bouncy ball bouncing up and down"\}} \\
\hline
``I have never felt this before.'' & {\ttfamily \{"keep": false, "reason": "No Sensation"\}} \\
\hline
``It feels sort of like a heartbeat.'' & {\ttfamily \{"keep": true, "rewritten\_text": "A heartbeat"\}} \\
\hline
``NA'' & {\ttfamily \{"keep": false, "reason": "N/A"\}} \\
\hline
``This sensation reminds of my phones ringtone vibration.'' & {\ttfamily \{"keep": true, "rewritten\_text": "A mobile phone ringtone vibration"\}} \\
\hline
``I dont think so'' & {\ttfamily \{"keep": false, "reason": "No Sensation"\}} \\
\hline
``A really strong buzz.'' & {\ttfamily \{"keep": true, "rewritten\_text": "A really strong buzz"\}} \\
\hline
\end{tabular}
\end{table*}